\documentclass{PoS}

\title{Early Pulsar Observations with LOFAR }

\ShortTitle{Early Pulsar Observations with LOFAR }

\author{\speaker{Jason Hessels}%
       \thanks{These results are presented on behalf of the LOFAR
         Transients Key Science Project.  Special thanks also go to
         G. Heald, J. Swinbank, J. Romein, M. Wise, and V. Gajjar for
         very important contributions to this work.  J. Hessels is a
         Veni Fellow of the Netherlands Organization for Scientific
         Research (NWO).}\\
       Netherlands Institute for Radio Astronomy (ASTRON), Dwingeloo, the Netherlands\\
       University of Amsterdam (UvA), Amsterdam, the Netherlands\\
       E-mail: \email{hessels@astron.nl}}

\author{Ben Stappers, Tom Hassall, Patrick Weltevrede\\
        University of Manchester\\}

\author{Anastasia Alexov, Thijs Coenen\\
        University of Amsterdam\\}

\author{Joeri van Leeuwen, Vlad Kondratiev, Jan David Mol\\
        Netherlands Institute for Radio Astronomy (ASTRON)\\}     

\author{Michael Kramer, Aris Noutsos\\
        Max Planck Institute for Radio Astronomy\\}

\author{Aris Karastergiou\\
        University of Oxford\\}

\abstract{This contribution to the proceedings of ``A New Golden Age
  for Radio Astronomy" is simply intended to give some of the
  highlights from pulsar observations with LOFAR at the time of its
  official opening: June 12th, 2010.  These observations illustrate
  that, though LOFAR is still under construction and astronomical
  commissioning, it is already starting to deliver on its promise to
  revolutionize radio astronomy in the low-frequency regime.  These
  observations also demonstrate how LOFAR has many ``next-generation"
  capabilities, such as wide-field multi-beaming, that will be vital
  to open a new Golden Age in radio astronomy through the Square
  Kilometer Array and its precursors.}

\FullConference{ISKAF2010 Science Meeting \\
   June 10 -14 2010\\
    Assen, the Netherlands}

\begin{document}

\section{Introduction}

LOFAR is a low-frequency radio interferometer that covers the
$10-240$\,MHz frequency range, the lowest 4 octaves of the ``radio
window" above the Earth's ionospheric cut-off at $\sim 10$\,MHz.  This
is a relatively unexplored range of the electromagnetic spectrum,
where there is still exciting potential for serendipitous astronomical
discoveries.  Ultimately there will be at least 40 LOFAR stations in
the Netherlands, along with at least 8 international stations in
Germany, England, France, and Sweden.  Each of these stations contains
48/96 Low-Band Antennas (LBAs), which cover the range $10-90$\,MHz, as
well as 48/96 High-Band Antennas (HBAs) covering the range
$110-240$\,MHz.  The relatively small frequency range occupied by FM
radio broadcasts is purposely filtered out to avoid this strong interference.  Each station is capable
of forming multiple ``station beams", which are then correlated and/or
summed as necessary in a central Blue Gene P (BG/P) supercomputer in
Groningen.  The product of station beams times total bandwidth is
48\,MHz per station, giving LOFAR a remarkably large fractional
bandwidth and field-of-view (FOV).  LOFAR will be the most powerful
and flexible low-frequency radio telescope ever built; arguably, it is
already.  It is also an important precursor to the Square Kilometer
Array (SKA), by demonstrating many relevant technologies for the first
time.  Roughly half of the LOFAR stations are currently complete, with
the remaining stations to be built in the next year.

Since the summer of 2007, the LOFAR Pulsar Working Group has been
performing observations with the existing LOFAR hardware in order to
commission the high-time-resolution ``beam-formed" modes necessary to
observe pulsars, ``fast transients", planets, the Sun, and flare stars
(to name a few sources).  A description of work on the ``Pulsar
Pipeline" is given in \cite{Alexov10}, also in these proceedings.  In
\S2, we describe some of the highlights of observational results
so-far achieved with LOFAR.  These observations nicely demonstrate
much of the functionality that will make LOFAR a powerful,
ground-breaking telescope for studying pulsars and other sources that
vary on sub-second time-scales.  Since the official opening of LOFAR
on June 12th, 2010, commissioning work in this area has only
intensified as we move closer to full operations.  We discuss the
future prospects for pulsars and LOFAR in \S3.  We note that a
comprehensive reference paper describing pulsar observing with LOFAR
is in preparation \cite{Stappers11}.

\section{Early Pulsar Observations}

\subsection{Wide-band Simultaneous Observations}

LOFAR's wide instantaneous frequency coverage in the $10-240$\,MHz
range is unique and can be exploited to study the frequency dependence
of pulsar emission, such as its spectrum, pulse-energy distribution,
and integrated pulse morphology.  Similar studies in the past have
suffered from very small available bandwidths, and the need to combine
non-contemporaneous data, e.g., \cite{Malov10}.

\begin{figure}[!h]
   \centering
   \includegraphics[width=3in]{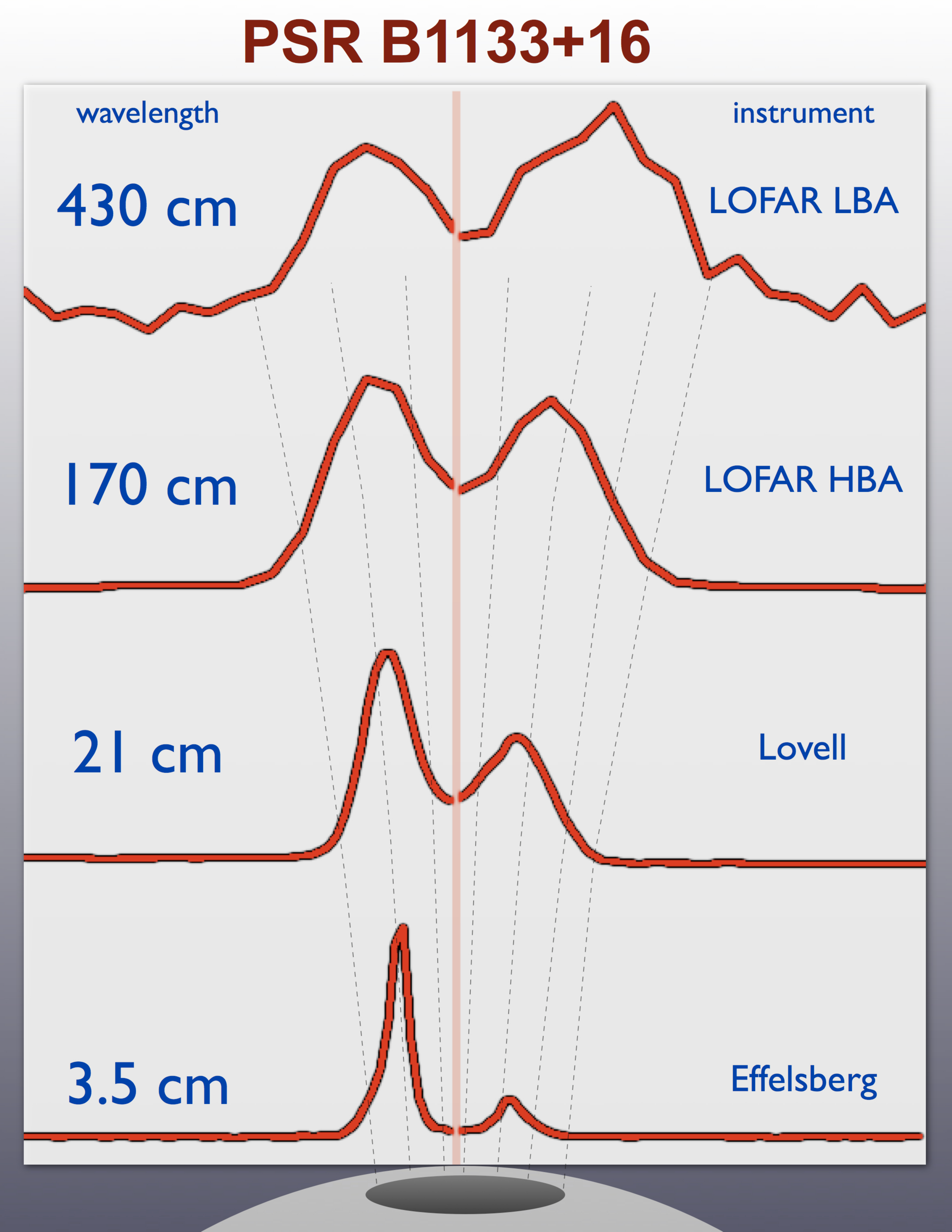}

   \caption{Simultaneous detection of pulses from PSR B1133+16
     in four widely spaced bands, using the Effelsberg telescope at
     3.5 cm wavelength (8.5\,GHz), the Lovell telescope at 21 cm
     wavelength (1.4\,GHz), and LOFAR HBAs and LBAs at 170 cm and 430
     cm wavelength (176\,MHz/70\,MHz), respectively. For each of the
     four bands, the figure shows the cumulative pulse intensity over
     this 1-hr observation as a function of rotational phase.  In the
     simplest model, the shape of the pulsar's pulsed emission maps
     the spreading of magnetic field lines above the pulsar's magnetic
     poles, also depicted schematically.}

   \label{fig:Wide}
\end{figure}

To demonstrate LOFAR's potential in this area, we performed a set of
simultaneous observations in which we observed 6 bright pulsars using
two LOFAR stations - one set to use the LBAs, the other set to use the
HBAs - in conjunction with the 74-m Lovell Telescope at Jodrell Bank
and the 100-m Effelsberg Telescope.  This unique combination of
telescopes allowed us to simultaneously observe the radio light from
these pulsars across wavelengths from only 3.5 centimeters (8.5\,GHz) up
to 7 meters (43\,MHz) - a factor of 200 difference.  This set a new
world record in simultaneous wavelength/frequency coverage for
pulsars.  Shown in Figure~\ref{fig:Wide} are the multi-frequency pulse
profiles for one of the pulsars, B1133+16.

In the future we plan to use sub-arrays\footnote{A sub-array is simply
  a collection of LOFAR stations configured to run in the same
  observing mode and pointing in the same direction.  It is possible
  to run multiple observations simultaneously such that groups of
  LOFAR stations can be running each in a separate mode, at the same
  time.} of the LOFAR stations in order to completely cover the
$10-240$-MHz frequency window simultaneously.  As each station can
provide up to 48\,MHz of instantaneous bandwidth, 5 sub-arrays are
required for complete spectral coverage.

\subsection{Millisecond Pulsars}

At low radio frequencies, interstellar scattering strongly affects the
maximum achievable time resolution of observations because multi-path
propagation of the signal causes it to be significantly smeared in
time.  The short ($\sim 100$\,$\mu$s) pulses from millisecond pulsars
are easily affected by this, but despite this it is still possible to
observe a large fraction of the known population of millisecond
pulsars using LOFAR \cite{Stappers08}.  As very few studies have been
done on the low-frequency emission properties of millisecond pulsars,
this is a potentially fruitful avenue of research.  Past studies have
indicated that millisecond pulsars may behave differently in this
frequency regime than their more slowly rotating,
higher-magnetic-field brethren (\cite{Stappers08} and references
therein).

\begin{figure}[!h]
   \centering
   \includegraphics[width=5in]{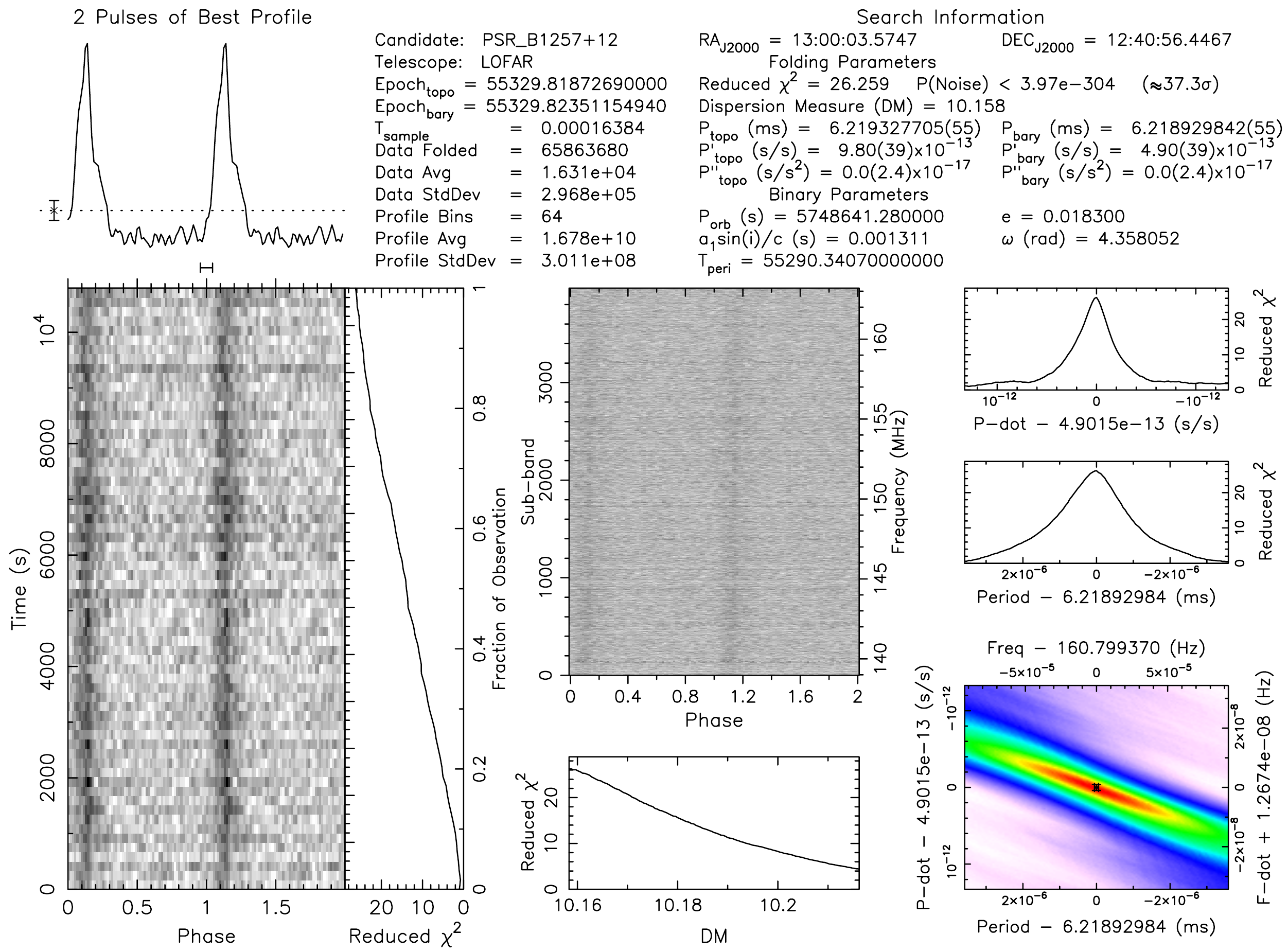}

   \caption{A diagnostic plot from the PRESTO pulsar software suite,
     which shows the clear detection of pulsations from the ``Planet
     Pulsar" B1257+12.  This pulsar is known to be orbited by at least
     2 planetary mass bodies - the first to be discovered outside of
     the solar system \cite{Wolszczan92}.  The left-most plot shows the pulse
     intensity as a function of observing time, with the cumulative
     pulse profile (repeated twice for clarity) shown at the top.  The
     central plot shows the signal strength across the 24\,MHz
     bandwidth, which was divided into $\sim 4000$ channels.  At
     center bottom, the signal to noise is shown as a function of
     trial dispersion measure.  Note how sensitive the signal strength
     is to the assumed dispersion measure.  At LOFAR observing
     frequencies, we will likely have to adjust the dispersion measure
     slightly on a per epoch basis because of small changes in our
     line-of-sight to the pulsar with time.  As can be seen here, the dispersion measure at the time of this observation deviates subtlely from the catalog value of 10.1655\,pc cm$^{-3}$.  The right hand sub-panels
     show an optimization of the signal strength on trial period and
     period derivative.  Observation meta-data and folding parameters
     are shown at the top of the plot.}

   \label{fig:MSP}
\end{figure}

We have detected a number of millisecond pulsars with LOFAR, including
the famous 6.2-ms ``Planet Pulsar" B1257+12.  Figure~\ref{fig:MSP} shows a diagnostic plot of this detection using the PRESTO software suite\footnote{http://www.cv.nrao.edu/$\sim$sransom/presto/}.
This was done using an incoherent dedispersion method with a large
number of spectral channels.  Soon we will be dedispersing these data
{\it coherently} in order to much better remove the dispersive effects
of the interstellar medium and to maintain the maximum time resolution
available in the system ($\sim 5$\,$\mu$s).

\subsection{Multi-Station Observations}

\begin{figure}[!h]
   \centering
   \includegraphics[width=3.5in]{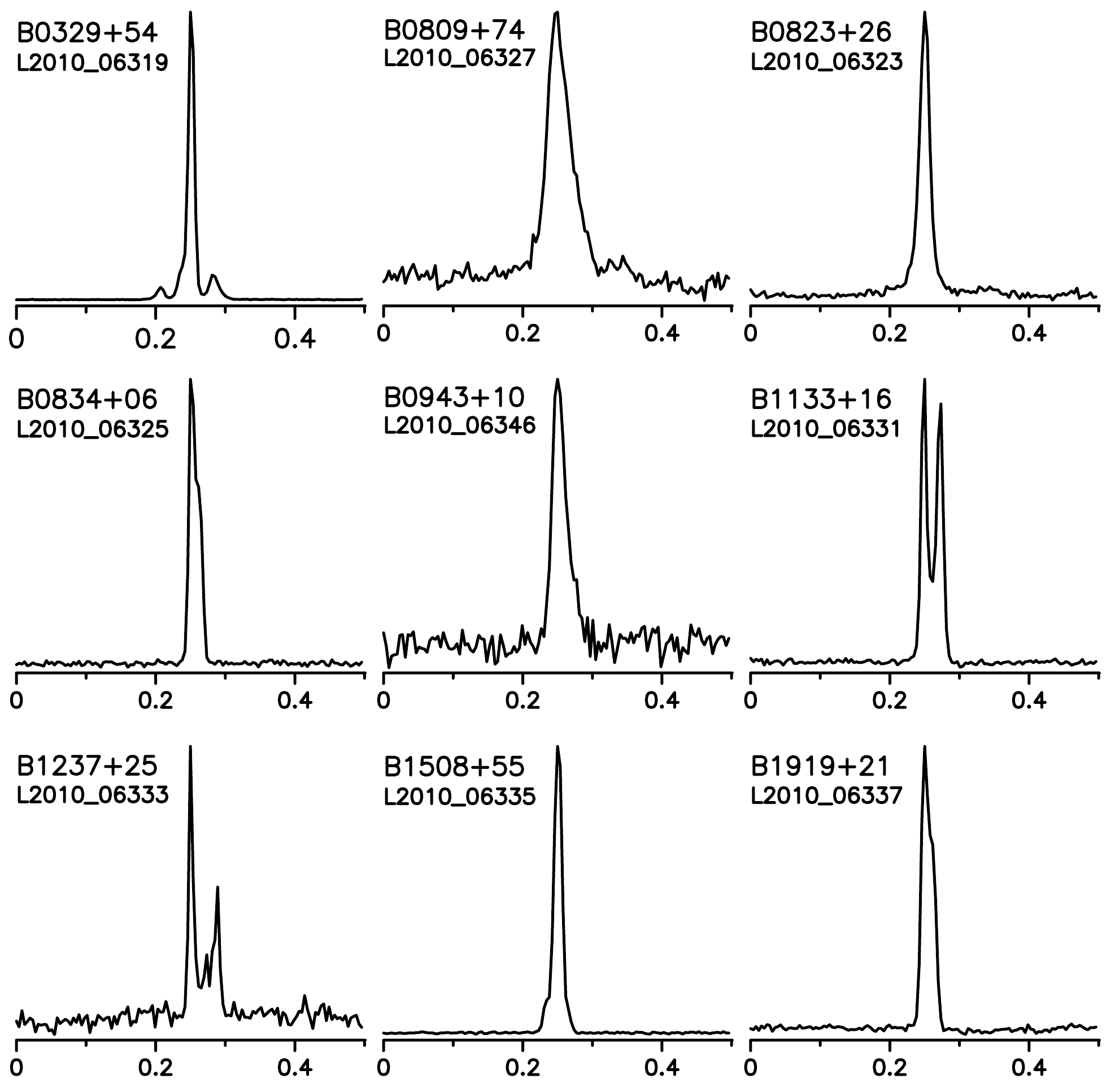} 

   \caption{Pictured are the average pulse profiles of nine pulsars,
     observed over the course of a weekend using an automatic
     scheduler.  Only half the rotational phase is plotted in each
     case in order to show finer features of the pulse
     morphology. Each pulsar was observed for roughly an hour with the
     HBA tiles (see accompanying observation ID), using different combinations of stations, with up to
     12 stations at a time added incoherently.}

   \label{fig:Profs}
\end{figure}

Pulsar observations are always sensitivity limited, and combining the
signals from many LOFAR stations is critical for improving
sensitivity.  Ideally we will combine the stations ``coherently", with
a proper phase calibration - taking care of geometrical, instrumental,
and environmental effects - to form small tied-array beams on the sky.
So far, most multi-station LOFAR pulsar observations have combined the
stations incoherently, by simply summing the station powers with the
appropriate geometrical delays.  Figure~\ref{fig:Profs} shows data
from the incoherent combination of 12 Dutch stations.

These are some of the highest quality data we have taken thus far. In
particular, we note that even in the densely populated Netherlands we
have been able to manage and excise the majority of the radio
frequency interference by blanking only about 5\% of the spectral
channels.  The future holds even more promise: with the full set of
calibrated LOFAR stations added coherently, we should be able to
achieve another factor of 10 increase in sensitivity compared with
these detections.

\subsection{Simultaneous Imaging and Pulsar Observations}

On April 9, 2010, we took a 12-hr observation in which we
simultaneously ran the imaging and pulsar observing modes on the LOFAR
BG/P correlator.  This 12-hour observation used the HBAs of 7 core and
3 remote stations from 130-178 MHz.

\begin{figure}[!h]
   \centering
   \includegraphics[width=6in]{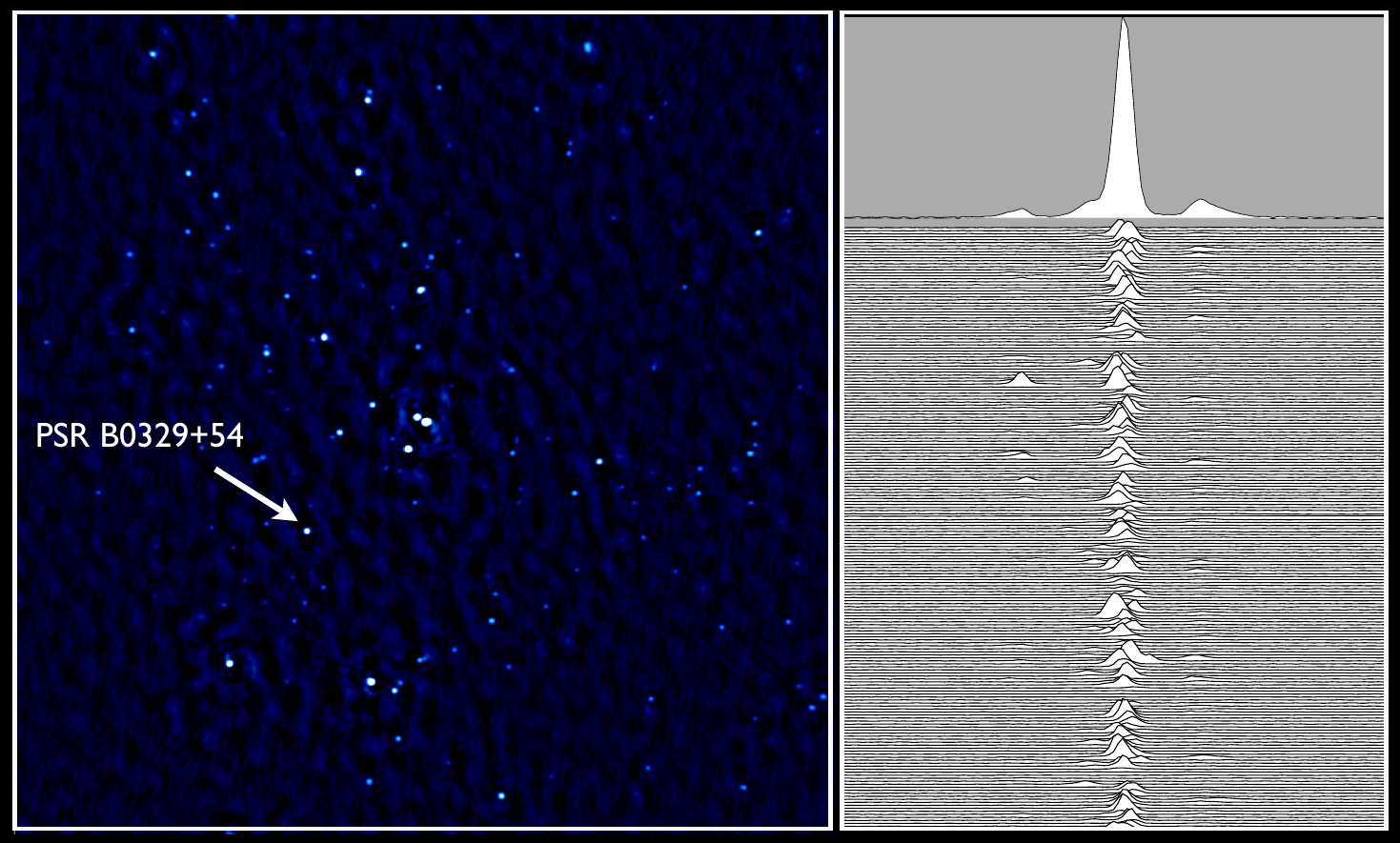}

   \caption{{\it Left:} A LOFAR HBA image of a $5^{\circ} \times
     5^{\circ}$ field containing the very bright pulsar B0329+54.
     This pulsar has a flux density of roughly 1\,Jy at 150\,MHz and
     is one of the brighter sources in the field.  {\it Right:} Pulse
     stack of roughly 200 pulses from the same imaging/beam-formed
     observation.  Significant pulse-to-pulse profile morphology and
     intensity variations are obvious.  The
     sum of these pulses is shown at the top.}

   \label{fig:Img_PSR}
\end{figure}

The imaging data were processed using the Standard Imaging Pipeline,
using the three brightest sources in the field for calibration, and
excluding short baselines. Following pipeline processing, an updated
sky model was derived from the image and used for a second pass of
calibration; the individual subbands were processed separately but
then combined before imaging into one large measurement set, so that
the deconvolution could be done properly. The Common Astronomy
Software Applications package (CASA\footnote{http://casa.nrao.edu/})
was used for the imaging.  The resulting combined image is shown in
Figure~\ref{fig:Img_PSR}.  The location of pulsar B0329+54 is
indicated by the arrow; the pulsar has a known flux of $\sim 1$\,Jy at these frequencies.  The full
field is very large: roughly $5^{\circ}$ across.

The simultaneously acquired, beam-formed, time series data, with a
time resolution\footnote{For ease of processing, these data were
  heavily downsampled in time online.  Much higher time resolution, up
  to $5.12$\,$\mu$s - limited by the 195.3-kHz subbands that are sent
  from the stations - is achievable with the current system.  In the
  future, an inverse poly-phase filter on the station subbands may
  allow much higher time resolution still (down to $10$\,ns), which
  will be used for detecting radio emission from the wake of cosmic
  rays entering the Earth's atmosphere.  Such high time resolution
  also has applications for the study of pulsar giant pulses, though
  scattering remains a significant limiting factor on the effective
  time resolution achievable at these observing frequencies.} of
1.3\,ms, was processed using the Known Pulsar Pipeline.  A small
subset of roughly 200 single pulses is shown in the ``pulse stack" on
the right of Figure~\ref{fig:Img_PSR}, with the corresponding
cumulative pulse profile at the top.

This ``piggy-backing" mode will be a key aspect of LOFAR operations,
as it optimizes the use of available telescope time.  It is
particularly interesting for monitoring the sky for rare ``fast
transient" events and will be used along with interferometric imaging
as part of the LOFAR ``Radio Sky Monitor", which will regularly
monitor the sky to detect transients on time-scales of milliseconds to
years \cite{Fender08}.

\subsection{Multiple Station Beams}

Unlike most conventional radio telescopes, LOFAR stations can look in
multiple directions at the same time by distributing the total
observing bandwidth between multiple ``station beams" (FOVs)
simultaneously. This makes it possible to monitor a large fraction of
the sky at once and to observe multiple known sources during one
observation.  As such, the prospects for all-sky surveys and
monitoring for rare transient events are very exciting.  For the LBAs,
a particular station can form beams anywhere in the visible sky; for
the HBAs a preceding level of analog beam-forming at the level of the
individual HBA tiles limits the forming of station beams to within the
roughly $20^{\circ}$-wide FOV of the tile beam, which itself can be
pointed in any desired direction.

\begin{figure}[!h]
   \centering
   \includegraphics[width=5in]{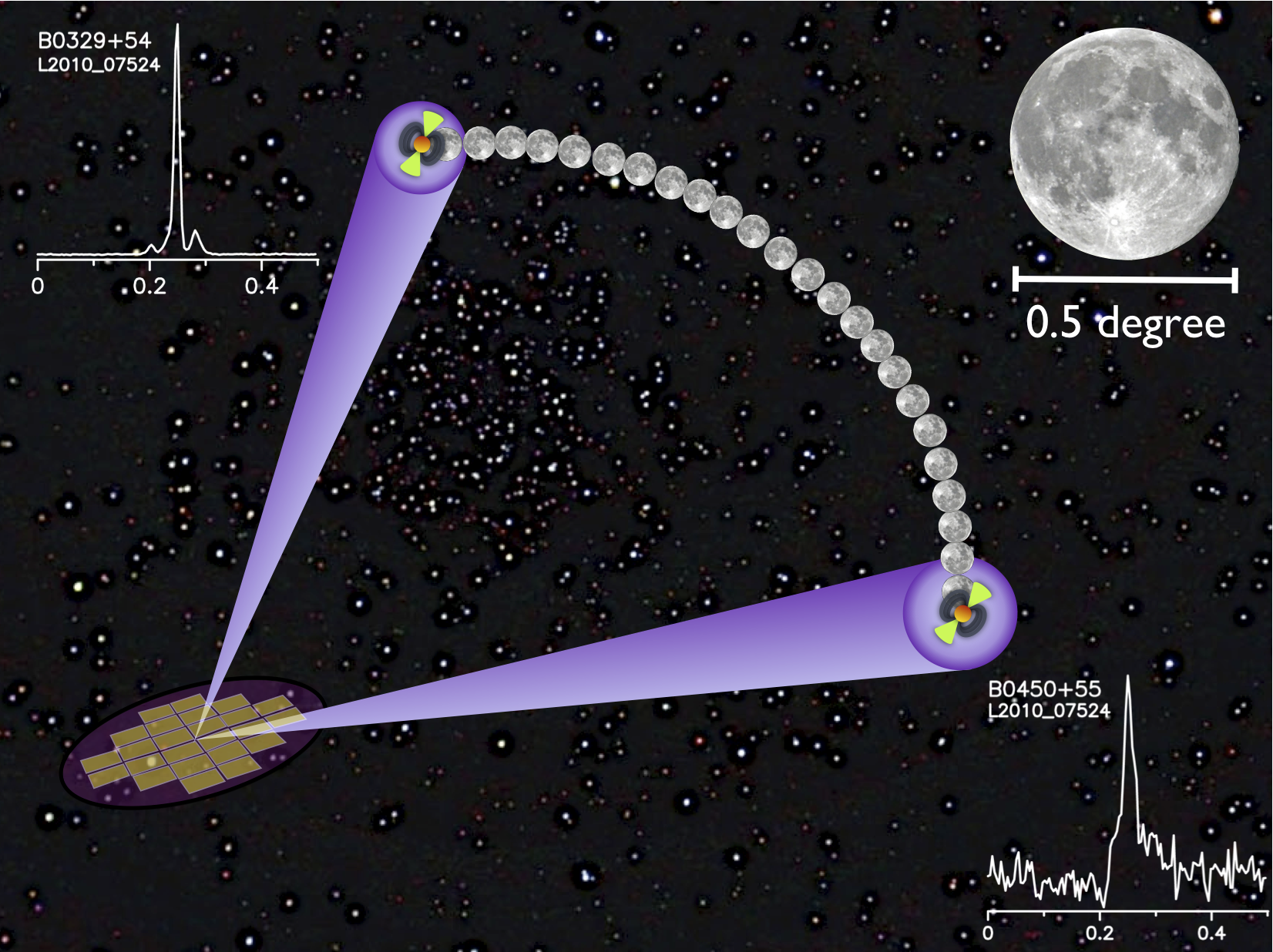}

   \caption{An artist's impression of the simultaneous detection of
     pulsars B0329+54 and B0450+55 using two station beams.  These
     pulsars are separated by close to $12^{\circ}$ on the sky,
     roughly 24 times the width of the full moon.}

   \label{fig:Multi_Beam}
\end{figure}

To demonstrate this impressive functionality, we observed two widely
separated pulsars simultaneously. Pulsars B0329+54 and B0450+55 are
separated by close to $12^{\circ}$ on the sky, or, to put it another
way, by roughly 24 times the width of the full moon
(Figure~\ref{fig:Multi_Beam}).  Pointing the 20$^{\circ}$ HBA tile
beam directly between the two pulsars, we formed two station beams and
centered one on each of the two pulsars. Tracking the sources for
30\,min, both the very bright B0329+54 (the brightest pulsar in the
northern sky) and the much fainter B0450+55 were clearly detected.

We plan to repeat this type of observation using more beams on more
sources. With the LBAs, each of which can effectively see the whole
sky above the horizon, it will be possible to form station beams that
are even more widely separated.  To do something similar with the HBAs
would require forming sub-arrays of the LOFAR stations, akin to a
``Fly's Eye" mode.

\subsection{Crab Giant Pulses}

\begin{figure}[htbp]
   \centering
   \includegraphics[width=4in]{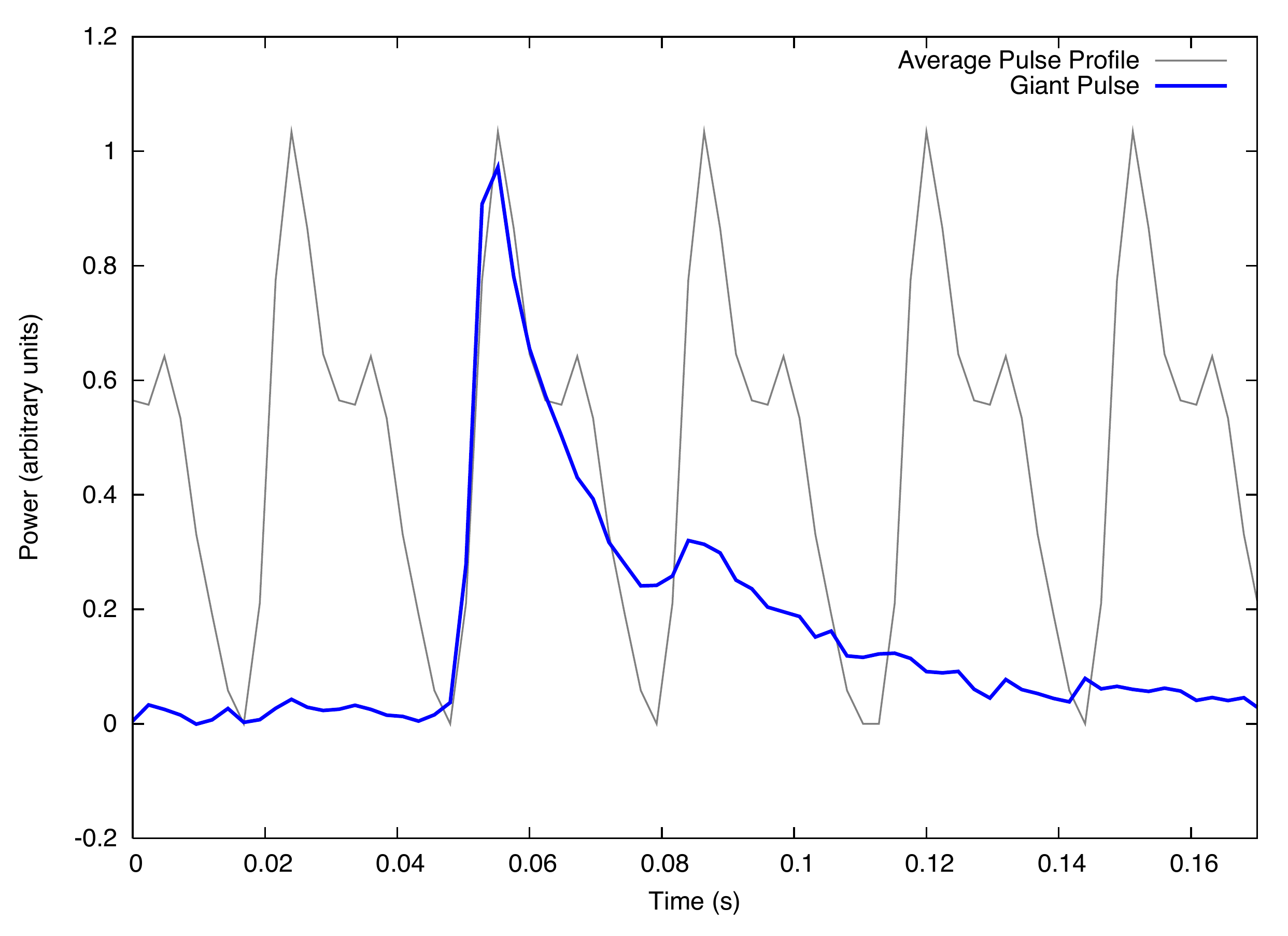} 

   \caption{Example giant pulse as observed with the LOFAR HBAs at a
     central frequency of 160\,MHz.  Interestingly, this particular
     pulse appears to be a double giant pulse, with two such pulses
     occurring separated by only one pulse period (this is not the interpulse).}

   \label{fig:Crab_Giant}
\end{figure}

The Crab pulsar is the proto-type of a small population of radio
pulsars known to emit short, extremely bright ``giant pulses".  These
pulses can intrinsically be as short as nanoseconds or less \cite{Hankins03}, 
and have the highest brightness temperatures of
any known astrophysical phenomenon.  At low radio frequency, the Crab
pulsar's giant pulses are severely scattered, creating a long
exponential tail on their trailing edge.  As an example,
Figure~\ref{fig:Crab_Giant} shows a heavily scattered giant pulse
detection with the LOFAR HBAs ($\nu_{center} \sim 160$\,MHz).  From
this figure, it can been seen that the length of the pulse is actually
scattered to a time-scale longer than the pulse period (the cumulative pulse-profile is shown in the
background).  Interestingly, this particular pulse appears to be a
double giant pulse, with two such pulses occurring separated by only
one pulse period (this is not the interpulse).  Similar detection techniques will be used to search for bright, dispersed pulses from extragalactic pulsars and even potential cosmological bursts.

In the future, observations of known sources like the Crab will
greatly benefit from the formation of tied-array beams (see \S3),
which will restrict the FOV and hence reduce the contribution of
background emission (in this case, the very bright Crab Nebula).

\subsection{Single Station ``Fringes"}

Pulsar observations are not only an excellent way to test LOFAR's
beam-formed modes; they are also well-suited for more generic
commissioning tests of the system.  As an example, we have taken good
advantage of the fact that the brightest pulsar in the northern
hemisphere, PSR B0329+54, fortuitously passes through the Zenith at
the latitude of the LOFAR core in Exloo, the Netherlands.  This has
permitted numerous tests that use the fact that the pulsar will drift
through the station beam of elements that are added with zero
geometrical delay.  This has been used, e.g., to successfully debug pointing and
tracking problems in the past.

\begin{figure}[!h]
   \centering
   \includegraphics[width=5in]{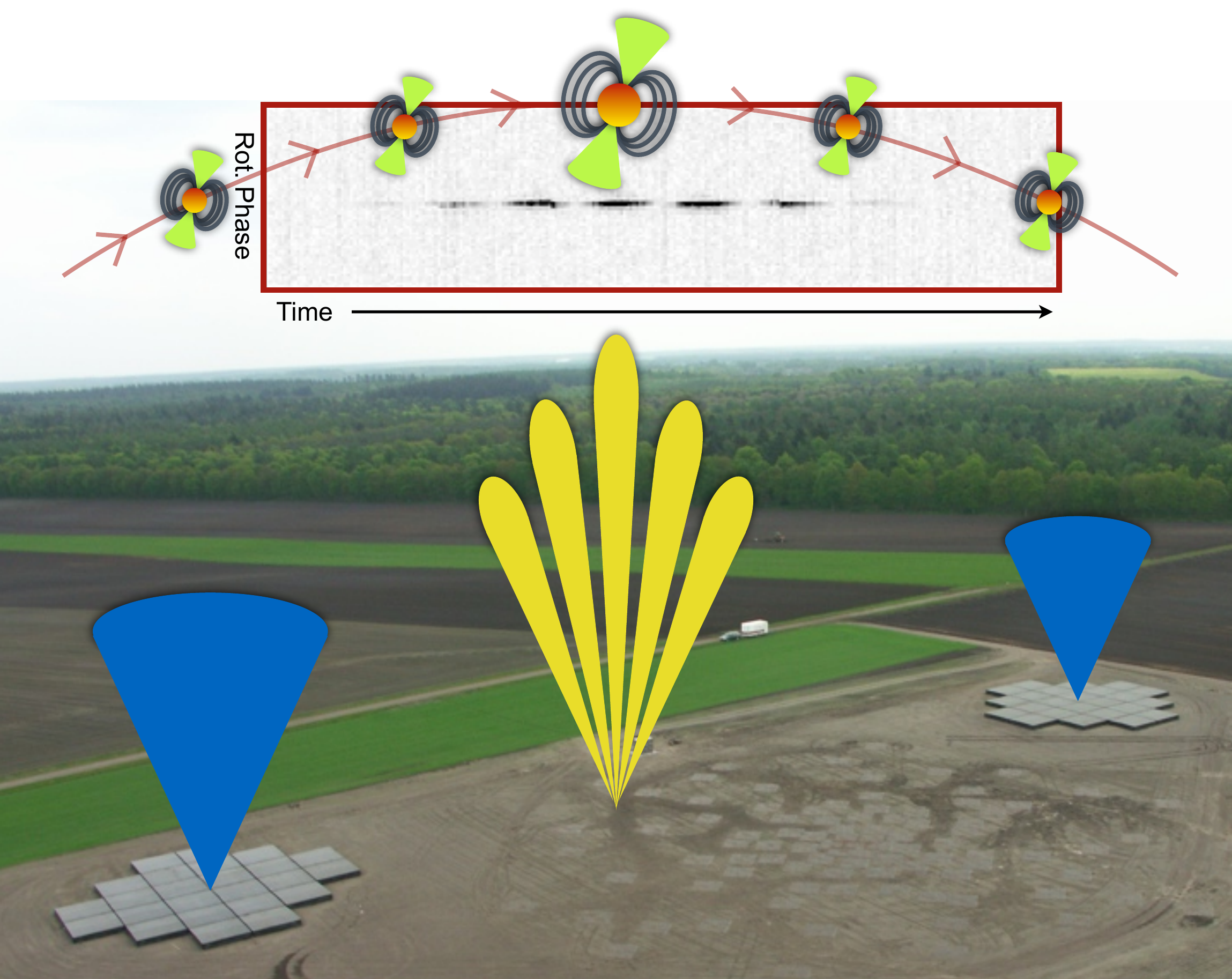} 

   \caption{Artist's impression of the very bright pulsar B0329+54 as
     it transits through the Zenith.  The red rectangle shows the
     actual data: the folded pulse profile as a function of rotational
     phase and observing time.  As the pulsar passes through the fan
     beams, it's observed strength is seen to rise and fall in
     accordance with the expected beam pattern.}

   \label{fig:Fringes}
\end{figure}

A particularly nice example is a transit observation in which PSR
B0329+54 passed through the summed beam of the two HBA ``ears" of core
station CS302.  Each core station has two sets of 24 HBA tiles
separated by roughly 130\,m in order to provide a wealth of short
baselines for imaging.  When combined coherently, the resulting,
instantaneous beam should form a fringe pattern of fan beams.  To test
that this was indeed the case, we observed a transit of PSR B0329+54
in this mode.  Figure~\ref{fig:Fringes} gives an artist's impression
of this setup and the red rectangle shows the folded pulse profile as
a function of rotational phase and observing time.  As the pulsar
passes through the fan beams, it's observed strength is seen to rise
and fall in accordance with the expected beam pattern.

\section{Conclusions and Future Prospects}

As shown here, commissioning of LOFAR for pulsar observations has 
made great strides in the last year.  The ability to simultaneously image 
and record at high time resolution coupled with LOFAR's proven multi-beaming 
capabilities will make it a powerful instrument for exploring the transient radio sky.

However, despite the many successful observations we have presented here, there
remains much commissioning work still to be done.  In fact, it is
likely that several years from now we will still be improving the
observational setup in order to reap the maximum benefit of LOFAR's
hardware.  In the meantime, however, we hope that increasingly
impressive observational results continue to be achieved.  Since the
official LOFAR opening, work on the beam-formed modes has continued in
earnest.  Here we comment on this current and future work.

{\bf Automated Observing: } The debugging of automated observing
pipelines, so that the modes that are already working can be used to
observe regularly and efficiently, is now well under way.  In late July
and early August 2010, we have done several all-weekend campaigns with
hundreds of $10-30$-min observations of known pulsars.  Currently the
down-time between successive observations is 3 minutes - still
impressive compared with a conventional telescope that is steered
mechanically - and is expected to improve.  These observations have
allowed us to clearly detect over 100 known pulsars; their wide-band, low-frequency pulse profiles
will be presented in a future paper.  These data also offer the
opportunity to exercise our proto-type search pipeline, which has
already been successful at detecting even faint, known pulsars and
with managing radio frequency interference excision.  An (expected)
consequence of LOFAR's complex beam pattern and sidelobes is that the
very brightest pulsars ($> 100$\,mJy at 150\,MHz) can be detected very
far from their true direction.  In the extreme case, we have even
detected PSR B1508+55 in a beam pointed 64$^{\circ}$ away from the pulsar's true
position.  This effect presents an additional challenge in pulsar surveys.

{\bf Coherent Dedispersion: } Especially for millisecond pulsars,
coherent dedispersion will be required.  It is possible to write out
the LOFAR station data as complex samples, which allows us to then
apply coherent dedispersion to the data offline.  A rudimentary
version of this pipeline is now working and the results are
encouraging.  Ultimately, it is our plan to implement this online on
the BG/P correlator so that observations of known pulsars can have the
benefit of coherent dedispersion without the burden of recording with
large data rates.

{\bf Tied-Array Beams: } To form proper tied-array beams requires a
precise calibration and correction of geometrical, instrumental, and
environmental phase delays between stations.  The geometrical delays
are of course the most straight-forward and are routinely applied when
we form incoherent summations of the station signals.  The spring and
summer of 2010 has seen much work dedicated to understanding and
correcting the instrumental phase delays between stations.  In
particular a new ``Single Clock" was installed near the site of the
LOFAR ``Superterp", which is the central core of the array, containing
the 6 innermost LOFAR stations.  This Single Clock provides a common
clock signal to these stations, which removes the need to calibrate
the clock drifts between stations.  It is possible that this could be
extended to other nearby stations in the LOFAR core.  Early tests of
the Single Clock have been very encouraging, and we now believe we are
close to combining all the 6 Superterp stations coherently.  As these
stations are within a circular diameter of only about 350\,m, there is
likely no need for further calibration of differential ionospheric
phase delays between the stations.  Proper correction for the ionosphere
remains the greatest challenge to forming tied-array beams with all
core stations, and we are currently investigating calibration schemes
that make use of LOFAR's ability to simultaneous image and take
pulsar-like data.

\end{document}